\documentclass[journal=nalefd,manuscript=letter]{achemso}
\usepackage[utf8]{inputenc}

\usepackage{color} 
\usepackage{xcolor}
\usepackage{amsfonts}
\usepackage{ulem}
\usepackage{hyperref}
\usepackage{amsmath}
\usepackage{siunitx}  
\usepackage{pgfplots}
\pgfplotsset{compat=1.18}

\definecolor{myPink}{RGB}{235, 80, 150}
\definecolor{myPurple}{RGB}{130, 70, 205}
\definecolor{myPin}{RGB}{200, 80, 150}
\definecolor{myPurpl}{RGB}{100, 60, 200}

\title{Tensor-driven geometric phase in nonlinear AlGaAs metasurfaces}
\keywords{AlGaAs, metasurfaces, geometric phase, second harmonic generation}

\author{Giorgio Guercio}
\author{Andrea Gerini}
\affiliation{Materiaux et Phénomènes Quantiques, Université Paris Cité and CNRS, 10 rue A. Domon et L. Duquet, 75013 Paris, France}
\author{Kristina Frizyuk}
\affiliation{Karlsruhe Institute of Technology, Kaiserstrasse, 12, Karlsruhe, 76131, Germany}
\author{Martina Morassi}
\author{Aristide Lemaître}
\affiliation{Centre de Nanosciences et de Nanotechnologies, Université Paris-Saclay and CNRS, 10 Boulevard Thomas Gobert, 91120 Palaiseau, France}
\author{Costantino De Angelis}
\affiliation{Department of Information Engineering, University of Brescia, Via Branze 38, 25123, Brescia, Italy}
\alsoaffiliation{INO-CNR, Via Branze 45, 25123, Brescia, Italy}
\author{Luca Carletti}
\affiliation{Department of Information Engineering, University of Brescia, Via Branze 38, 25123, Brescia, Italy}
\alsoaffiliation{INO-CNR, Via Branze 45, 25123, Brescia, Italy}
\author{Giuseppe Leo}
\affiliation{Materiaux et Phénomènes Quantiques, Université Paris Cité and CNRS, 10 rue A. Domon et L. Duquet, 75013 Paris, France}
\alsoaffiliation{Institut universitaire de France}
\email{giuseppe.leo@u-paris.fr}

\begin{document}

\begin{abstract}
    Dielectric metasurfaces provide a unique platform for efficient harmonic generation
    and optical wavefront manipulation at the nanoscale.
    While several approaches are available for performing wavefront shaping, the one exploiting geometric phase streamlines significantly the design and fabrication process.
    It has been recently shown that, in III-V semiconductor alloys, the  rotation  of  the crystal axes affects the phase and amplitude of second-harmonic generation (SHG) induced by circularly polarized light~\cite{Carletti_2024}.
    Based on this notion, we fabricated and characterized two aluminum gallium arsenide metasurfaces displaying the versatility of the geometric phase design approach through nonlinear beam steering and structured-light generation on the harmonic field.
    
\end{abstract}

\section{Introduction}

The emergence of optical metasurfaces has introduced a powerful planar approach for engineering ultrathin optical devices with flexible and tunable control over light at subwavelength resolutions~\cite{Yu_2014, Yu_2011, Lalanne_1998, Hsiao_2017, Kildishev_2013}.

By employing carefully designed nanostructures, called meta-atoms, metasurfaces allow precise control over the phase, amplitude and polarization of light in a highly compact and unique manner that surpasses the capabilities offered by standard refractive optics.
Metasurfaces can tailor the wavefront of incident light using geometric (
which includes Pancharatnam-Berry, PB) phase~\cite{Tymchenko_2015}, guidance~\cite{Lalanne_2002}, or resonant-phase~\cite{Decker_2015} (Huygens) approaches.

Guidance along nanopillars has long been known for providing high diffraction efficiencies ($\approx$ 80\% routinely) at large deviation angles~\cite{Lalanne_2002, Khorasaninejad_2016,Lalanne_1999} and being resilient to changes of the illumination direction~\cite{Lee_2000}. 
However, the guidance approach requires high aspect ratios, complicating the fabrication of such devices. 
The Huygens approach~\cite{Decker_2015, Yu_2015, Shalaev_2015, Lin_2014, Pfeiffer_2013} can be understood from a simple picture where every meta-atom provides 
electric and magnetic dipolar resonances~\cite{Decker_2015, Staude_2013} with opposite phases and achieve forward scattering~\cite{Kerker_1983}.
Together, these resonances make every meta-atom behave as a secondary source emitting purely forward-propagating elementary waves, hence the name of Huygens’ metasurfaces~\cite{Decker_2015} that is often given to this resonant approach. 
This design strategy allows for lower aspect ratios when compared to the guidance approach, but it often requires numerous different elements to completely fill the $0-2\pi$ phase profile needed in arbitrary wavefront shaping.

PB-metasurfaces overcome these technological constraints by streamlining the wavefront engineering via simple adjustments of the linear phase of the transmitted or reflected light through localized nanostructure rotations~\cite{Yu_2014, Georgi_2021, He_2019, Ding_2014}.

Beside their advantages in the linear regime, metasurfaces emerged a few years ago as an exciting and potentially useful technological platform for nonlinear optics~\cite{Maxim2014,Carletti:15,Krasnok_2018, Zubyuk_2021, Vabishchevich_2023}.
This is due to their great potential to reduce the size of devices while adding exotic functionalities.
The ease of fabrication introduced by PB-metasurfaces allows for the realization of simple yet versatile nonlinear nanophotonic devices; relevant examples include nonlinear plasmonic metasurfaces and photon-spin dependent nonlinear geometric-phase structures.\cite{Huang_2018, Li_2015, Keren_Zur_2016}

Additionally, hybrid metasurfaces that combine plasmonic architectures with semiconductor quantum wells have been employed to achieve efficient control of nonlinear wavefronts.\cite{Tymchenko_2015} 
Recent studies have further explored the selective response of dielectric nanostructures in nonlinear harmonic generation processes.\cite{Liu_2020, Reineke_Matsudo_2022,Liu_2023} 

All-dielectric metasurfaces, in particular, offer a compelling alternative to their plasmonic counterparts by exploiting Mie resonances to reduce dissipative losses while improving robustness and efficiency, thereby making them attractive platforms for optical wavefront manipulation. 

Nevertheless, these works typically consider only materials whose nonlinear response remains unaffected by in-plane rotations, which severely limits the range of available materials, accessible degrees of freedom, and the achievable generation efficiency.

It has been recently demonstrated how the rotation of the crystal axes of a thin film of (100) {AlGaAs} affects the phase and amplitude of second-harmonic generation ({SHG}) induced by circularly polarized light of opposite chirality.\cite{Carletti_2024} 
While such semiconductor is isotropic from a linear point of view, the observed phase stems from the anisotropy of the second-order nonlinear tensor, impacting solely the second harmonic (SH) signal. Thus, this phenomenon represents a purely nonlinear geometric phase effect. 
In metasurface design, combining an anisotropic second-order nonlinear tensor that is sensitive to in-plane rotations with rotated resonators enables access to a novel phase function that has not been previously reported. 
In this work, we demonstrate experimentally the capabilities of this phase function, by designing, fabricating and characterizing two metasurfaces: one for nonlinear beam steering, and the other for the generation of structured harmonic light.

\section{Design}
We start our analysis by choosing a resonator geometry possessing in-plane broken symmetry to allow SH emission in the out-of-plane vertical direction.
As demonstrated in a previous work~\cite{Carletti_2024}, (100) {AlGaAs} shaped as a half-cylinder effectively breaks the 
symmetry, ensuring that the radiation of SH along the the vertical direction is allowed. 
The allowed total angular momentum (TAM) projections for SHG are $m^{2\omega} = 2m^{\omega} + m_{\chi} + \mathfrak{n}\nu$, where $m^{2\omega}$ and $m^{\omega}$ are the TAM projections of the SH and pump light respectively, $m_{\chi} = \pm2$ is the additional momentum provided by the symmetry of the nonlinear second-order AlGaAs tensor, $\mathfrak{n}$ is the $\mathfrak{n}$-fold rotational symmetry of the structure, and $\nu \in \mathbb{Z}$~\cite{Nikitina2023}. 
For the selected structure, $\mathfrak{n} = 1$, resulting in $m^{2\omega} = 2m^{\omega}\pm2 + \nu$. 
Therefore, all integer values of $m^{2\omega}$ are admixed, including $m^{2\omega} = \pm1$ which corresponds to modes with circularly polarized electric fields that have non-zero radiation along the vertical direction.
\begin{figure}
  \includegraphics[width = 0.8\textwidth]{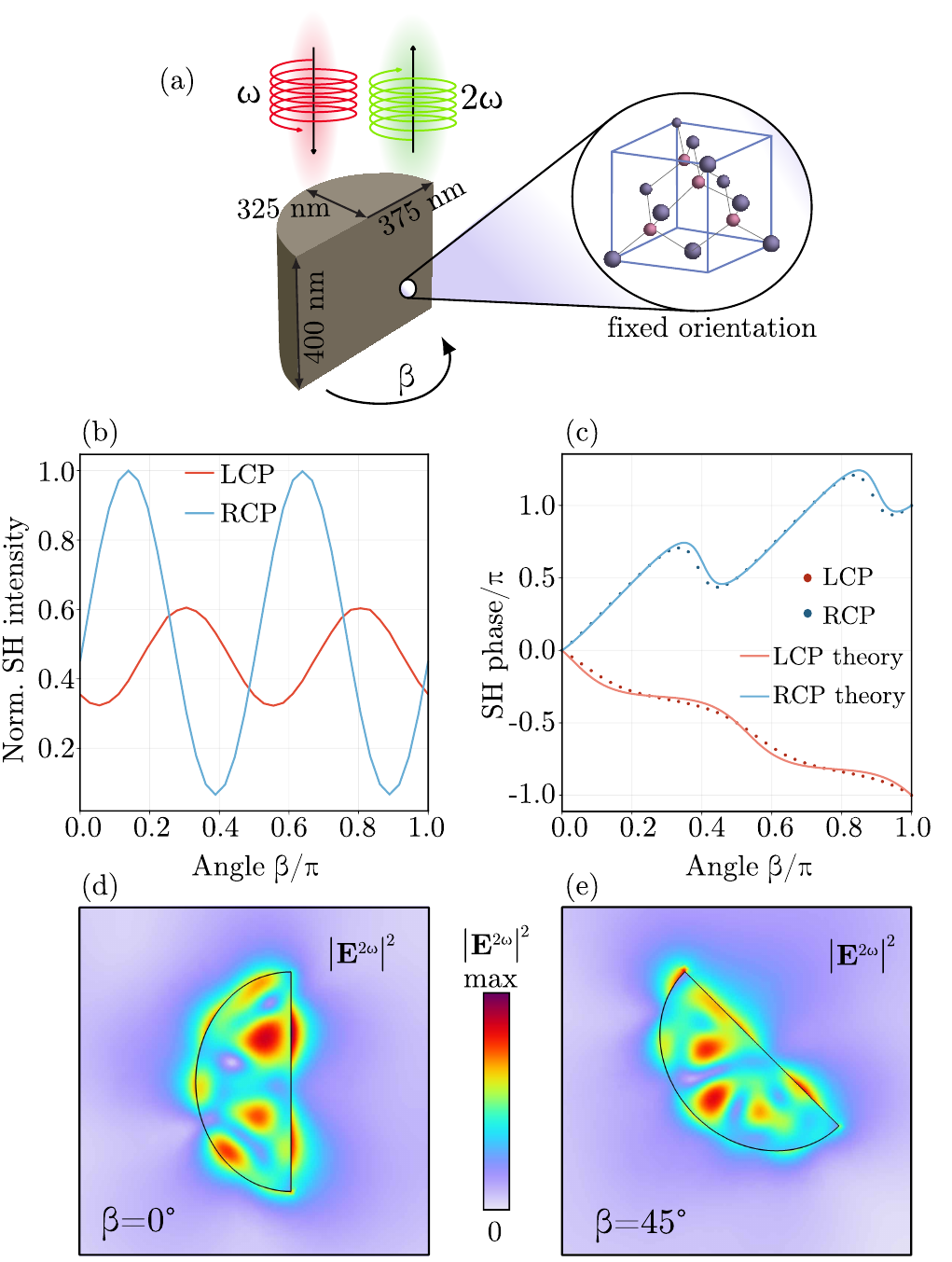}
  \caption{(a) Schematic of the AlGaAs nanoantenna geometry, with $\beta$ the in-plane rotation angle. The circularly polarized pump and SH beams are shown in red and green, respectively. (b) Normalized SH intensity emitted in the vertical direction (i.e., orthogonal to the substrate plane) versus $\beta$. (c) Phase of the SH electric field emitted in the vertical direction versus $\beta$. (d,e) SH electric field amplitude in the plane bisecting the nanoantenna for $\beta=\qty{0}{\degree}$, (d), and $\beta=\qty{45}{\degree}$, (e).}
  \label{fgr:D1}
\end{figure}

To show that the phase of the generated SH light is controlled by the rotation of the resonator, we study the SHG by a single AlGaAs resonator as a function of the orientation of the nanoantenna with respect to the Cartesian reference coordinate system [Fig.~\ref{fgr:D1}(a)]. 
Throughout this work, only the resonator undergoes in-plane rotation, while the orientation of the AlGaAs crystalline axis remains fixed and identical to that of the reference structures.
The SHG phenomenon is modeled using fully vectorial electromagnetic computations based on the finite element method, implemented through the commercial software COMSOL Multiphysics. 
The resonator height is \qty{400}{\nano\meter} and the elliptical semi-axes are \qty{325}{\nano\meter} and \qty{375}{\nano\meter} as shown in Fig. \ref{fgr:D1}(a).
The structure is illuminated with light at a wavelength of \qty{1550}{\nano\meter} and right circularly polarized (RCP) electric field.
Figures~\ref{fgr:D1}(b) and (c) show the normalized intensity and the phase of the SH generated in the same direction as the pump beam, for left-circularly polarized (LCP) and RCP electric field.
We can observe that the SH intensity varies as a function of the angular orientation of the resonator [Fig.~\ref{fgr:D1}(b)]. 
This is due to a variation of the orientation of the internal electric field with respect to the AlGaAs crystal axes as shown in Figs.~\ref{fgr:D1}(d) and (e). 
The SH phase, shown in Fig.~\ref{fgr:D1}(c), exhibits a dependence on the rotation angle, as discussed later in the text.
Namely, for a RCP pump, the phase of RCP SH acquires a phase of $\beta$, whereas the phase of LCP SH acquires a phase of $-\beta$. 
These results show that the phase of the SH depends on the geometrical rotation of the structure and is only weakly modulated by resonance effects. 

This phenomenon can be understood by considering TAM conservation together with the rotation properties of the nonlinear tensor.  
Let us begin by examining a rotation of the entire system by an angle $\beta$. 
By definition of TAM, such a rotation introduces phase shifts of $-\beta m^{\omega}$ and $-\beta m^{2\omega}$ for the pump and SH fields, respectively.

If the material’s nonlinear response is invariant under rotations by $\beta$, the only difference between the original and the rotated structure lies in the phase of the field coupled into the resonator. 
By introducing a time shift to the pump field, the two situations can be mapped onto one another. 
To establish the phase relation between the original and rotated configurations, we therefore resynchronize the pump fields by adding a temporal delay $\Delta t = -{m^\omega \beta}/{\omega}$,  
which imparts an additional phase factor of $2m^\omega\beta$ to the SH field. 
Combining this with the TAM-induced phase of the rotated system yields  
\begin{align}
    \varphi(\beta) = -m^{2\omega}\beta + 2m^\omega\beta .
    \label{eq:result_a}
\end{align}
This result applies whenever rotation does not affect the nonlinear susceptibility tensor. 
For instance, for $m^\omega = 1$ and $m^{2\omega} = \pm 1$, we recover the well-known relations $\varphi(\beta) = \beta$ and $\varphi(\beta) = 3\beta$.

However, Eq.~\ref{eq:result_a} is no longer valid when the nonlinear tensor itself transforms under rotations, as is the case for AlGaAs. 
For zincblende materials, the nonlinear susceptibility transforms under rotation around the $z$-axis in the same way as $\sin(2\phi)$,
where $\phi$ is the azimuthal coordinate. 
This means that rotation by $\beta$ therefore produces a new effective tensor:
\begin{equation}
\label{eq:tensrot}
    \chi^{(2)\prime} = \cos(2\beta)\chi^{(2)}_1 + \sin(2\beta)\chi^{(2)}_2 ,
\end{equation}
where $\chi^{(2)}_2$ corresponds to $\chi^{(2)}_1$ rotated by $\pi/4$, and $\chi^{(2)}_1$ is taken to be aligned with the crystallographic axis ($x\parallel [100]$).

As described in Refs.~\cite{menshikov2025near, Nikitina2024}, the SH field is then a linear superposition of the fields generated by each tensor component, provided that only the nonlinear tensor (and not the structure itself) is rotated. 
If we imagine rotating the crystal rather than the nanostructure, the SH field becomes  
\begin{equation}
    \mathbf{E}^{2\omega}= \cos(2\beta)\mathbf{E}_1 + \sin(2\beta)\mathbf{E}_2 ,
\end{equation}
where $\mathbf{E}_1$ and $\mathbf{E}_2$ correspond to the fields generated by the two tensor orientations but differ in phase because their coordinate frames are rotated. 
Each field is expanded into a series that contains not only the two polarization components with $m=\pm 1$ but also contributions from other TAM projections.
For a chosen polarization component, the corresponding expansion coefficients denoted $a$ and $b$ are the complex amplitudes multiplying that specific component, and their values are set by the resonant response of the nanostructure.
The resulting SH intensity for a particular component is therefore  
\begin{equation}
    I(\beta) \sim \left| a\cos(2\beta) + b\sin(2\beta) \right|^2 ,
\end{equation}
where the values of the coefficients $a$ and $b$ may be obtained comparing with the numerical calculation results at $\beta = 0$ and $\beta = \pi/4$.

To determine the SH phase of a nanoresonator rotated by $\beta$ in a material where the nonlinear tensor is itself rotation-sensitive, we combine the tensor-induced contribution with the TAM-derived terms:  
\begin{align}
     \varphi(\beta) &= \varphi_\chi(\beta) - m^{2\omega}\beta + 2m^\omega\beta ,
     \label{eq:alles_ist_hier} \\
     \varphi_\chi(\beta) &= \arg\!\left(a\cos(2\beta) + b\sin(2\beta)\right)
     \label{eq:2beta}
\end{align}
where $\varphi_\chi(\beta)$ is the additional phase contribution to Eq. \ref{eq:result_a} due to the nonlinear tensor transformation. Here, $a$ and $b$ generally differ for the LCP- and RCP-generated SH components.

We apply Eq.~\ref{eq:alles_ist_hier} to our scenario and the results are overlapped with the numerical results in Fig.~\ref{fgr:D1}(c). In particular, the model let us clearly observe that the SH phase is governed by two distinct contributions:  
(i) the geometric phase arising from rotation of the nanoresonator, yielding the familiar $\beta$ and $3\beta$ dependencies, and  
(ii) the nonlinear-tensor contribution, which introduces a $2\beta$-dependent phase term for both LCP and RCP SH fields (Eq. \ref{eq:2beta}). 
The same approach applies to other materials, with the specific dependence determined by the rotation of the $\chi^{(2)}$ tensor.

\begin{figure}
  \includegraphics[width = 0.9\textwidth]{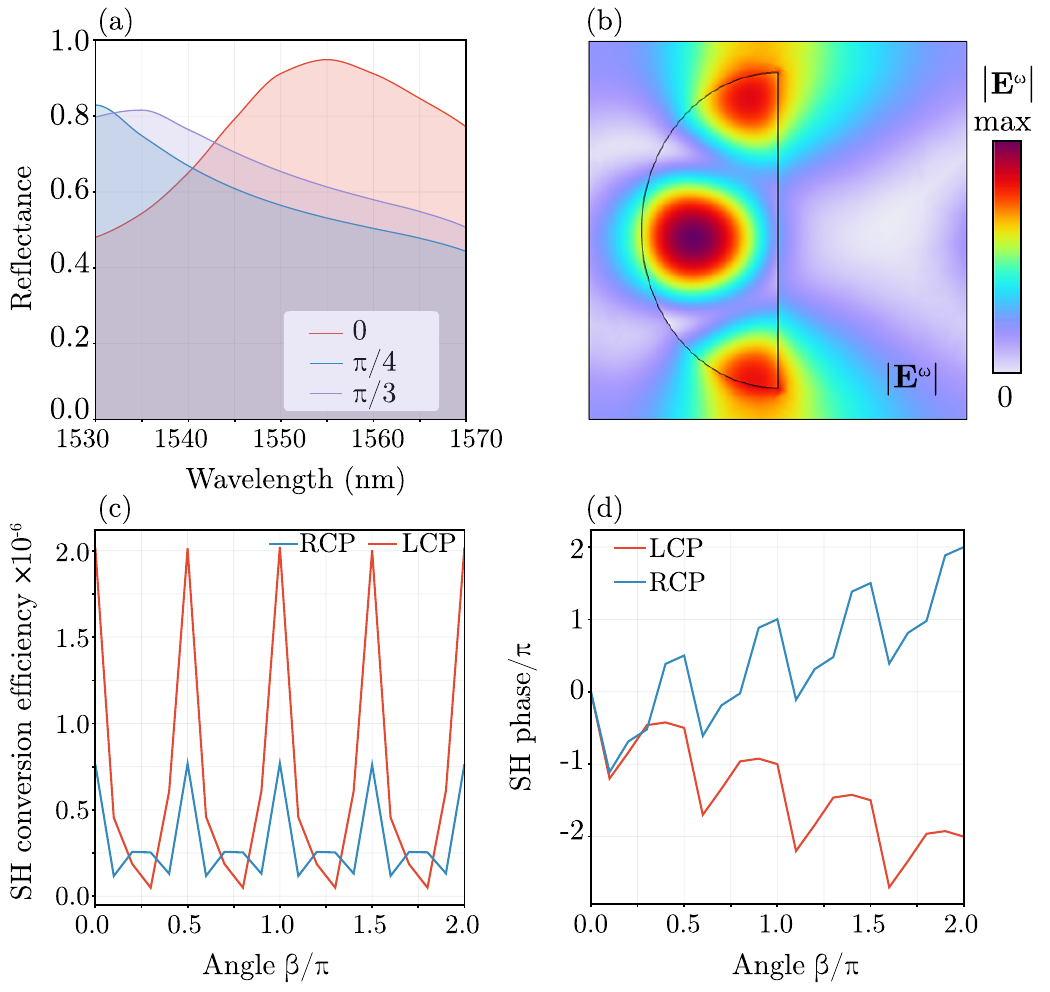}
  \caption{(a) Numerically calculated reflectance as a function of wavelength and meta-atom rotation angle $\beta$ for periodic array of resonators with period \qty{900}{\nm}. (b) Electric field amplitude in the $ xy$-plane bisecting the resonator. (c) normalized SH conversion efficiency and (d) phase as a function of the meta atom rotation angle $\beta$.}
  \label{fgr:D2}
\end{figure}
Having confirmed the capability to control the phase of the emitted SH in a nonlinear geometric-phase approach, we designed a resonant metasurface at a wavelength of \qty{1550}{\nano\meter}. 
The dimensions of the unit cell are the same of those of the nanoantenna of Fig.~\ref{fgr:D1}(a). 
The periodicity of the metasurface is set at \qty{900}{\nano\meter} to obtain an optical resonance at the pump wavelength.
The reflectance of the structure, calculated with COMSOL and reported in Fig.~\ref{fgr:D2}(a), shows a broad peak centered at a wavelength of \qty{1555}{\nano\meter}. 
Figure~\ref{fgr:D2}(b) shows the pump's electric field at a wavelength of \qty{1550}{\nm}, evaluated in the $xy$-plane that bisects the resonator, from regular metasurfaces, \textit{i.e.}, comprising unit cells with identical orientation angles. 
Figures~\ref{fgr:D2}(c) and (d) report the conversion efficiency and phase of the SH emitted in the same direction of the pump and in the fundamental diffraction order (0,0). 
As it can be seen in Fig.~\ref{fgr:D2}(d), the SH phase follows the expected trends which is overlapped to a periodic modulation. 
Similarly, the same modulation is observed in the SH intensity [Fig.~\ref{fgr:D2}(c)]. This modulation is due to the shift in the resonant wavelengths as a function of the orientation of the unit cell, as it can be seen for three representative cases of \qtylist[parse-numbers = false]{0;\pi/4; \pi/3}{\radian} in Fig.~\ref{fgr:D2}(a).

Such effect can, in principle, be attenuated by choosing a design that is more tolerant to rotations of the unit cell and in experiments using a pulsed pump beam that has a finite spectral bandwidth.

In the last step of the design of nonlinear wavefront shaping AlGaAs metasurfaces, we used Fig. \ref{fgr:D2}(c) to build a look-up table. In particular, we chose four elements corresponding to the rotation angles \qtylist[parse-numbers = false]{0;\pi/2;\pi;3\pi/2}{\radian}, and designed two metasurfaces: one for beam steering and one for generation of structured light. 

\section{Fabrication}
The sample has been fabricated on a molecular beam epitaxy consisting of: (100) {GaAs} substrate,  \SI{90}{\nano\meter} transition layer GaAs-to-Al$_{0.98}$Ga$_{0.02}$As, \SI{2000}{\nano\meter} Al$_{0.98}$Ga$_{0.02}$As, and \SI{400}{\nano\meter} Al$_{0.18}$Ga$_{0.82}$As.

The \SI{100}{\micro\meter}-diameter metasurfaces have been defined using e-beam lithography (EBL, Raith Pioneer II, \SI{20}{\kilo\volt} acceleration voltage) on \SI{100}{\nano\meter}-thick layer Hydrogen SilsesQuioxane resist (HSQ, Applied Quantum Materials Inc., H-SiOx-15 6\% in Buytl Acetate) whose adhesion was improved by spin coating the promoter SurPass 3000 (Micro Resist Technology). The sample has then been developed in diluted AZ400K (1:3 in deionised water, MicroChemicals GmbH) for \SI{1}{\minute}.
The pattern was then transferred to the Al$_{0.18}$Ga$_{0.82}$As layer using an inductively coupled plasma reactive ion etching (ICP-RIE) with a {SiCl$_4$/Ar} gas mixture (\SI{2}{sccm}/\SI{30}{sccm}, Sentech SI500).
Finally, the sample was selectively oxidised at \SI{390}{\celsius} for \SI{1}{\hour} \SI{45}{\minute} in order to have the Al$_{0.18}$Ga$_{0.82}$As metasurfaces on the low-refractive index AlOx layer (approximately \SI{1.6}{\micro\meter} thick).

\section{Measurements}
\subsection{Beam steering metasurface}
\begin{figure}
  \includegraphics[scale = 0.9]{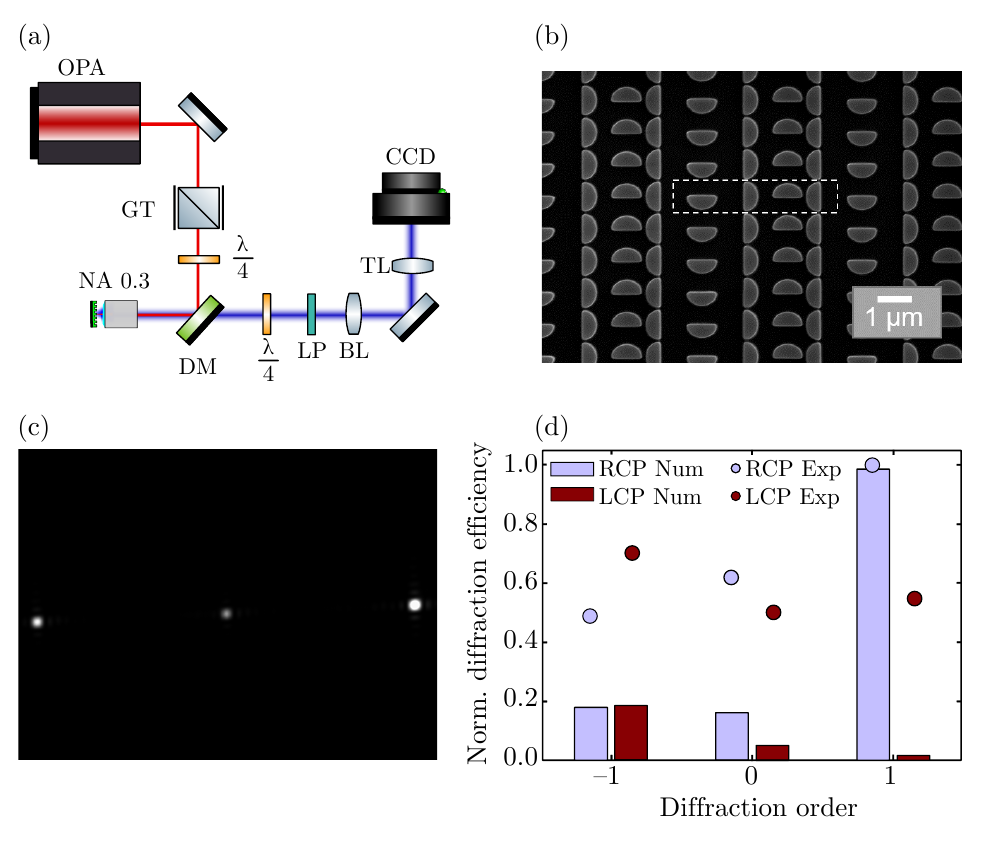}
  \caption{(a) Experimental setup. (b) SEM image of the beam steering metasurface. Metasurface supercell highlighted with dashed rectangle. (c) Fourier space image of the metasurface SH signal. (d) Polarization analysis of the various diffraction orders.}
  \label{fgr:BS}
\end{figure}

The first measured sample is a beam steering nonlinear metasurface. The design consists in a uniform array of a single supercell made of four rotated elements, as shown in Fig.~\ref{fgr:BS}(b). The figure shows a SEM zoom-in on the metasurface, with the dashed lines highlighting the supercell area. 
The element disposition allows to obtain a SH phase profile of the following type: $\varphi(x) = {\pm 2\pi x}/{P}$, where \qty[parse-numbers = false]{2\pi}{\radian} is the total phase shift along the supercell, $P=N\cdot D$ is the supercell dimension along $x$, $N$ is the number of resonators per supercell, and $D = \qty{900}{\nm}$ is the resonator spacing. 
For $N=4$, a deflection angle of $\pm$\qty{12}{\degree} with respect to the normal is obtained for the emitted SH in reflection depending on the polarization configuration. 
The measurement has been carried out with a nonlinear microscope used in reflection, as shown in Fig.~\ref{fgr:BS}(a). 
A Glan-Taylor polarizer and $\lambda/4$ waveplate turn the linearly polarized $\lambda_\mathrm{FF}=\SI{1550}{\nano\meter}$ beam coming from the \qty{200}{\femto\second} OPA into right circularly polarized (RCP) light. Since our sample works in reflection (due to the presence of the AlOx/GaAs substrate), a dichroic mirror reflects the pump light, while it transmits the SH light. 
A 0.3 numerical aperture (NA) objective has been employed for focusing on the sample and collecting the SH generated by the metasurface. 
A removable Bertrand lens allows to image the Fourier plane on a CCD camera.
Fig.~\ref{fgr:BS}(c) shows the appearance of three equispaced spots on the camera. 
Since both RCP and LCP SH are simultaneously generated by the metasurface, polarization-resolved measurements of the Fourier plane have been performed. 
A $\lambda/4$ waveplate and a linear polarizer have been placed after the dichroic mirror to measure the SH RCP and LCP contributions to each diffraction order. 
As derived from the lookup table in Fig.~\ref{fgr:D2}(c), the design directs RCP SH light to an angle of \qty{12}{\degree} and LCP SH light to \qty{-12}{\degree}, corresponding to the (+1,0) and (-1,0) diffraction orders, respectively. 
Figure~\ref{fgr:BS}(d) presents the measured SH intensity, depicted as dots in the histogram, categorized by polarization and diffraction order. 
The data reveal that, consistent with the metasurface design, RCP SH light is predominantly emitted in the (+1,0) order, while LCP SH light is mainly emitted in the (-1,0) order. 
This behavior is corroborated by fully vectorial numerical simulations of the beam steering metasurface design, shown as bars in Fig.~\ref{fgr:BS}(d).

\subsection{Structured light generating metasurface}
\begin{figure}
  \includegraphics[scale = 0.9]{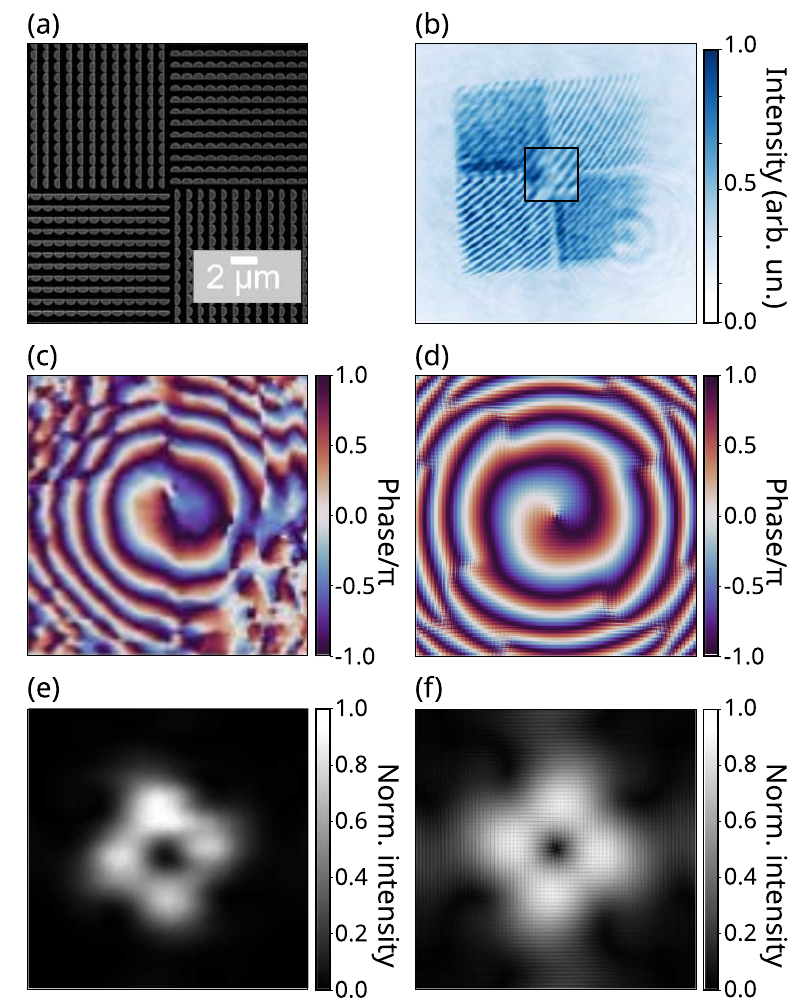}
  \caption{(a) SEM image of the structured light metasurface. (b) Near-field SH interference pattern. Inset shows a zoom-in of the fork-like pattern associated to a topological charge +1. (c) Measured and (d) simulated phase profile of the SH field emitted from the structured light metasurface. (e) Measured and (f) simulated far-field intensity map of the SH emitted by the metasurface.}
  \label{fgr:OAM}
\end{figure}

As a second demonstrator for the possibility to perform nonlinear wavefront shaping based on nonlinear-geometric phase AlGaAs metasurfaces, we measured a metasurface that generates optical vortices with unitary topological charge. 
The metasurface design is shown in Fig.~\ref{fgr:OAM}(a): four quadrants with identical resonators rotated by \qty[parse-numbers = false]{\pi/2}{\radian}. 
According to the lookup table, moving on a closed circular path centered at the common vertex of the four quadrants will result in a phase change of \qty[parse-numbers = false]{2\pi}{\radian} for RCP SH and \qty[parse-numbers = false]{-2\pi}{\radian} for LCP SH.

Successful encoding of the SH wavefront is first analyzed with a characterization of the phase texture of the SH beam. 
Experimentally, we let the SH beam from the device interfere with a reference beam in a generalized Mach-Zehnder configuration. 
The resulting fork-like interference pattern is shown in the zoomed inset in Fig.~\ref{fgr:OAM}(b). 
This interferogram not only demonstrates that the metasurface produces a vortex beam, but it also provides the value of the topological charge, $\ell$, carried by the SH beam. 
The value of $\ell$ can be estimated directly from the experimental interference pattern in Fig.~\ref{fgr:OAM}(b) by counting the number of bright fringes around the dislocation and it results $\ell=1$ for the measured metasurface. 
Alternatively, a Fourier transform of the interference pattern provides the added phases of the two beams that interfered. 
The result of this method is given in Fig.~\ref{fgr:OAM}(c), where a helical wave front with $\ell$ branches is clearly evidenced. 
Here $\ell$ can be defined as the closed-path integral of the phase gradient around the singularity $\ell = \frac{1}{2\pi}\oint\nabla\varphi \ \text{d}a$.
This phase profile aligns well with the one obtained from numerical calculations using the angular spectrum method, which qualitatively reproduces the propagation of SH light, as shown in Fig.~\ref{fgr:OAM}(d).
Non-idealities arise from the discretization of the phase mask,\cite{Jack_2008} and this is even more evident from the Fourier plane imaging of the metasurface, as shown in Fig.~\ref{fgr:OAM}(e): instead of a uniform bright ring shape, four bright spots appear. This behavior is confirmed by a simulation of an equivalent metasurface, as demonstrated in Fig.~\ref{fgr:OAM}(f).  

\subsection{Discussion}
\begin{figure}
  \includegraphics[scale = 0.9]{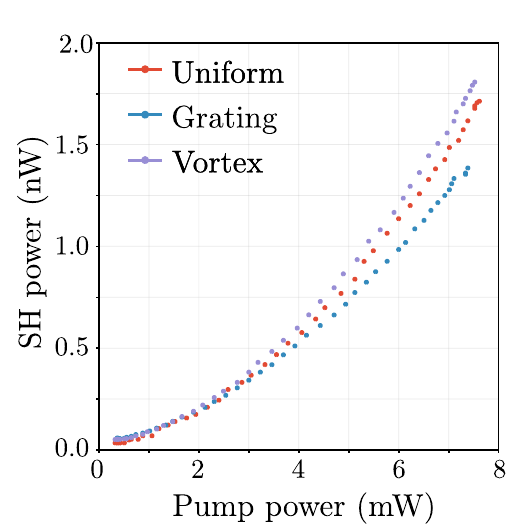}
  \caption{SH power as a function of pump power for different phase profiles.}
  \label{fgr:Eff}
\end{figure}

Our results demonstrate the capability of our framework for realizing nonlinear wavefront-shaping metasurfaces based on {AlGaAs} or, more generally, on materials with a zincblende crystal structure. 
The proposed approach is inherently broadband, as it does not rely on tuning optical resonances. 
Moreover, it does not compromise the performance of the final devices, including the conversion efficiency. 
To verify this, we estimated the conversion efficiency of the wavefront-shaping metasurfaces by comparing them with a uniform metasurface composed of identically tilted metaatoms.

The measurements were performed using a Si photodiode connected to a lock-in amplifier to suppress environmental noise. 
Figure~\ref{fgr:Eff} shows the radiated SH power as a function of the pump power for the three metasurfaces. 
We observe that the SH yield is comparable for the structured light and the uniform metasurface. 
The beam-steering metasurface exhibits a slightly lower power, likely due to partial radiation into higher diffraction orders that are not collected in the measurement because of the NA of the collection objective.

Finally, we note that the term geometric phase is used in various ways throughout the current literature. 
Canonical examples such as the Berry phase, the Aharonov–Bohm phase, and related phenomena are true holonomies~\cite{PhysRevLett.51.2167, Nakahara, RevModPhys.94.031001, Cayssol2021Apr} (see SI for further discussion). 
In contrast, the phase appearing in our work and in related metasurface studies does not exhibit a clear holonomic interpretation (or at least we have not identified one). 
This discrepancy introduces ambiguity and raises questions regarding consistent terminology.
One possible way to reconcile these usages is to adopt a broader definition in which geometric refers to any technique that controls the phase response through structural geometry rather than through wavelength- or material-dependent mechanisms. 
In conclusion, we demonstrated a design principle that exploits the interplay between the {AlGaAs} nonlinear tensor and geometric-phase control, enabling consistent {SH} efficiency across different metasurface designs while maintaining a straightforward fabrication process. 
We validated this approach through the experimental characterization of two nonlinear wavefront-shaping metasurfaces implementing fundamental functionalities as beam steering and structured-light generation. 
The framework is inherently broadband and can be readily adapted to metasurfaces supporting either local or nonlocal optical resonances. 
Overall, our results expand the set of available strategies for engineering nanophotonic devices capable of generating tailored light fields, with potential impact in areas such as telecommunications, quantum optics, sensing, and optical manipulation.

\begin{acknowledgement}
 K.F. gratefully acknowledges support from the Alexander von Humboldt Foundation. L.C. acknowledge partial financial support from the European
Union’s under Next Generation EU- PNRR - M4C2, investimento 1.1 - PRIN 2022 - NoLimiTHz (id:
2022BC5BW5) CUP D53D23001140001. G.L. Acknowledges ANR-DFG funding (Projet ANR-22-CE92-0090 MEGAPHONE).

\end{acknowledgement}

\begin{suppinfo}
    
\end{suppinfo}

\section{Author contributions}
G.G. carried out the experimental characterization, A.G fabricated the sample, K.F. formulated the theory and provided analytical insight, L.C. performed the simulations and modelling; M.M. and A.L. provided the epitaxy; L.C., C.D.A, and G.L. supervised the project.
G.G. and A.G. wrote the first draft, all authors contributed to the discussion of the results, the manuscript preparation, and its revision. L.C. conceived the original idea.

\bibliography{bibliography}

\end{document}


\maketitle

\begin{abstract}
This Supplementary Information provides an extended theoretical discussion of the results presented in the main text. 
\end{abstract}

\medskip

\smallskip

\tableofcontents

\section{Eigenmodes of the nanostructure}
\begin{figure}
  \begin{center}
  \includegraphics[width=0.5\textwidth]{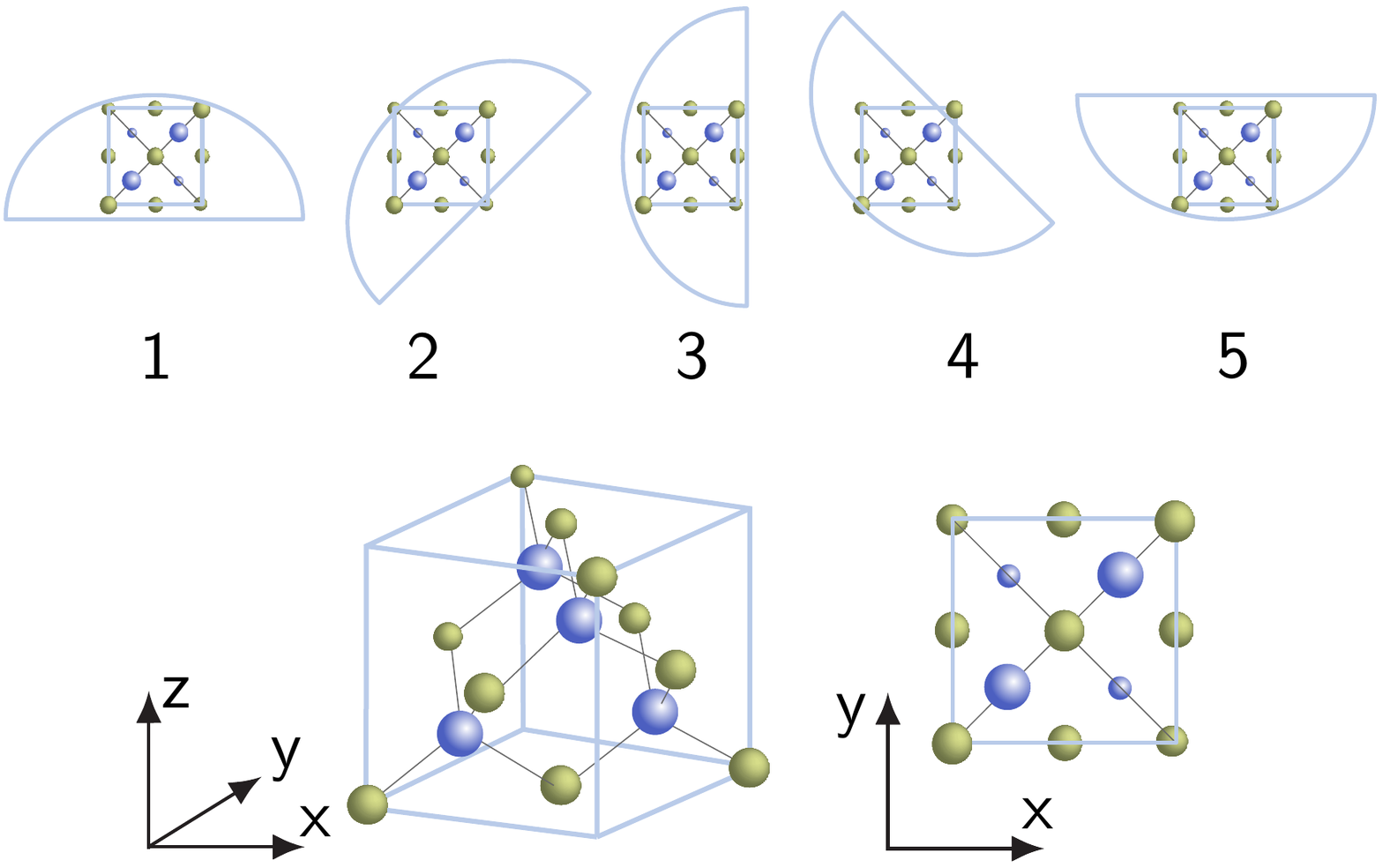}
  \caption{Relative orientation of the crystalline lattice and the structure.
  Configurations 1, 3, and 5 are equivalent up to a sign change of the $\chi^{(2)}$ tensor, and configurations 2 and 4 are also mutually equivalent up to this sign.
  Therefore, the amplitudes of the excited eigenmodes at $2\omega$ coincide for the structure rotated by $\pi/2$, due to the same coupling process.}
  \label{fgr:lattice}
  \end{center}
\end{figure}
We will first fix the orientation of the nanostructure as depicted as number~1 in Fig.~\ref{fgr:lattice}.
The semi-disk structure has very low symmetry, with only one reflection plane, $x=0$.
This means that there are only two types of eigenmodes: even and odd ones with respect to this reflection~\cite{Gladyshev_Frizyuk_Bogdanov_2020,Xiong_Xiong_Yang_Yang_Chen_Wang_Xu_Xu_Xu_Liu_2020}.
We denote them as follows:
\begin{equation}
   \text{even:  }\vb E_{ex} \sim \vb E_{ex, ey} + \vb E_{ex, oy},  
   \label{eq:evenmode}
\end{equation}
\begin{equation}
   \text{ odd:   }\vb E_{ox} \sim \vb E_{ox, ey} + \vb E_{ox, oy}.
   \label{eq:oddmode}
\end{equation}
We divided the fields of these eigenmodes into two parts, one of them is even with respect to the reflection in the $y=0$ plane, and the other is odd.
For the orientation 5 in Fig.~\ref{fgr:lattice} the fields will have the opposite form:
\begin{equation}
   \text{even:  }\vb E_{ex} \sim \vb E_{ex, ey} - \vb E_{ex, oy}, 
   \label{eq:evenmode5}
\end{equation}
\begin{equation}
   \text{ odd:   }\vb E_{ox} \sim \vb E_{ox, ey} - \vb E_{ox, oy}.
   \label{eq:oddmode5}
\end{equation}
One can write the corresponding expressions for all the orientations presented, using this language. For the orientations 2 and 4 one should exploit parities under reflection in planes $x=\pm y$.
\section{Theory of SHG phase dependence}
\subsection{Restricted set of rotation angles}
\begin{figure}
  \begin{center}
  \includegraphics[width=0.5\textwidth]{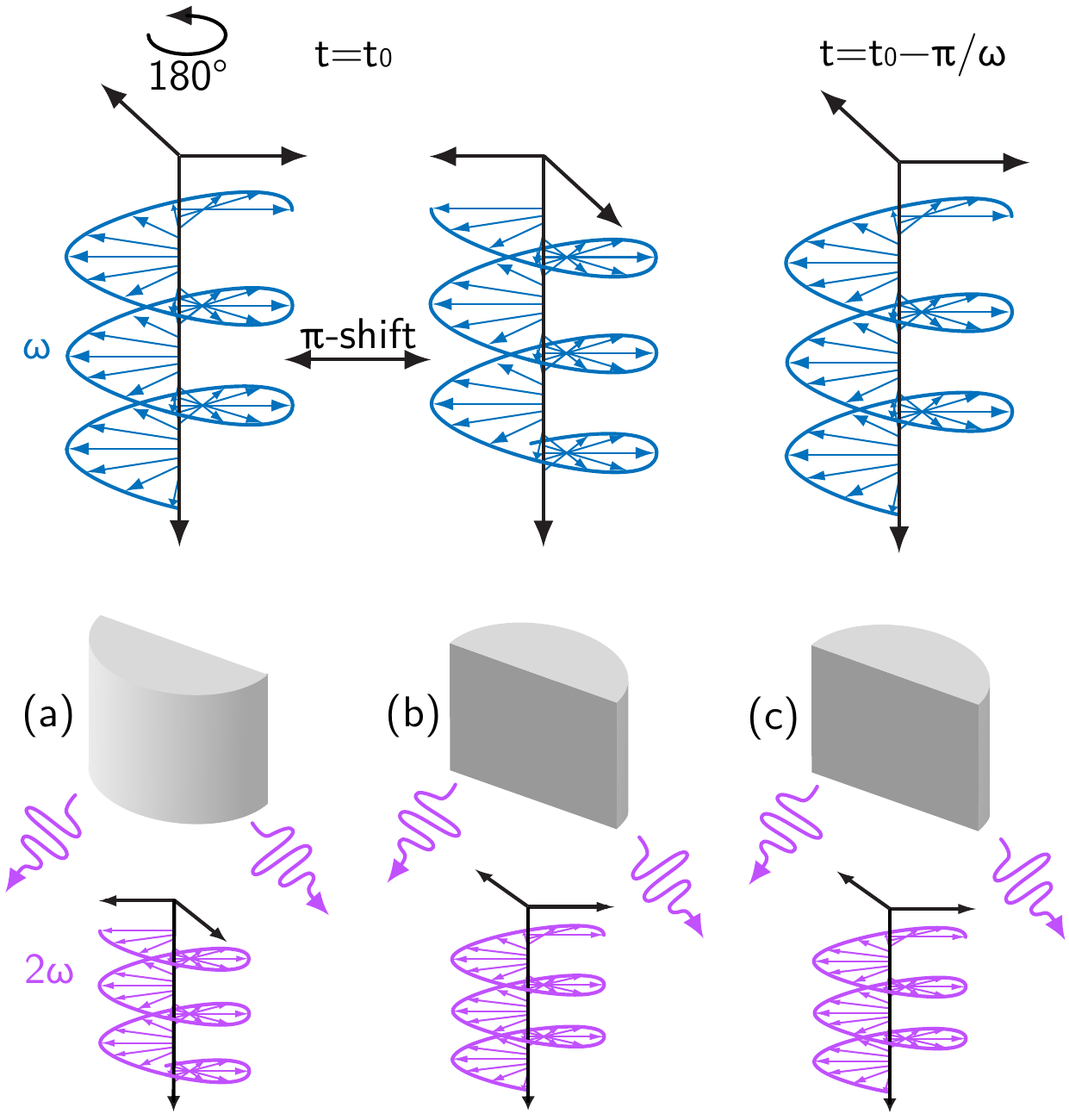}
  \caption{Illustration of the approach for a particular case. 
  The nanostructure is irradiated by a circularly polarized plane wave, and the SH contribution with the same polarization is considered. 
  (a) Reference system at $t_0$. 
  (b) $\pi$-rotated system at $t_0$. 
  (c) $\pi$-rotated system at a shifted time such that the incidence coincides with that in (a).}
  \label{fgr:theory}
  \end{center}
\end{figure}
In this section, we describe in detail an approach to finding the SHG phase for a rotation $\beta$ of the nanostructure that depends only on the TAM projections and the frequencies of the input and output waves, assuming that a few additional conditions are met.
This approach is similar to that described earlier~\cite{Liu_2023, Tymchenko_2015, Li_2015}, but is considered from a slightly different perspective, treating the nanostructure as a black box whose specific internal properties are not essential.
This allows the approach to be more easily generalized to any structure, any incident and generated TAM projections, and to higher-order harmonics.
The conditions that should be satisfied for the approach to work for a rotation angle $\beta$ without any numerical simulations are:
\begin{enumerate}
    \item  The generated TAM projection should be allowed by the selection rules for the harmonic generation process~\cite{Frizyuk2019, Nikitina2024};
    \item The nonlinear susceptibility tensor is invariant or only changes sign.
\end{enumerate}
First, let's assume that the tensor is invariant (it doesn't necessarily mean that the lattice is invariant~\cite{Nikitina2024}).
For AlGaAs half-cylinder nanoresonators, this corresponds to a 180 degrees rotation (see Fig.~\ref{fgr:theory}).
We are only interested in the SH contributions  with $m^{2\omega} = \pm 1$.
The TAM projection of the incident wave is denoted as $m^{\omega}$, which equals $1$ 
in our case.
Symmetrically, they behave just like circularly polarized plane waves, which is depicted in Fig.~\ref{fgr:theory} for simplicity.
Our approach suggests comparing Fig.~\ref{fgr:theory}(a) and (b), where (b) is simply (a) rotated by $\pi$ as a whole, together with the incident and generated waves.
If we look at panel (b) at the time $\omega t = \omega t_0 - \pi$ [Fig.~\ref{fgr:theory}(c)], where $t_0$ is the time instant used in panels (a) and (b), we see that the incident wave coincides with panel (a). 
The generated wave, however, remains identical to panel (b) at $t_0$, because it advances by $2\pi$ due to its doubled frequency.
Thus, the SH signal is rotated by $\pi$ when comparing panels (a) and (c), which have the same incidence but a rotated structure.
The same reasoning can be applied to the case of a $90^\circ$ rotation of the structure, but with an additional phase of $\pi$, because under a $90^\circ$ rotation the $\chi^{(2)}$ tensor changes sign~\cite{Nikitina2024}.
The general algorithm for obtaining the phase of the SH component with TAM projection $m^{2\omega}$ under a rotation by $\beta$ is as follows:
\begin{enumerate}
    \item Rotate the system as a whole by $\beta$. 
    The incident wave acquires a phase $-m^\omega \beta$, and the generated wave acquires a phase $-m^{2\omega}\beta$, according to the definition of TAM~\cite{video1}.
    \item Shift the time such that the incident wave has its initial phase, i.e., $\Delta t = -{m^\omega\beta}/\omega$.
    The generated wave then acquires an additional phase $2m^\omega\beta$ due to the doubled frequency.
    \item If the tensor changes sign under a rotation by $\beta$, add a phase of $\pi$ to the generated wave. 
    The final phase is
    \begin{equation}
      \varphi(\beta) =  -m^{2\omega}\beta + 2m^\omega\beta  \ (\ +\pi) 
    \end{equation}
    \item For harmonic generation of order $N$, replace the $2m^\omega\beta$ term with $Nm^\omega\beta$.
\end{enumerate}
In our case, $m^\omega = 1$ and $m^{2\omega} = \pm 1$, which provides the required phases.

\subsection{Theory of the SHG for an arbitrary angle $\beta$}
\begin{figure}[h!]
    \centering
\begin{tikzpicture}
  \pgfmathsetmacro{\aone}{0.354842}
  \pgfmathsetmacro{\rone}{0.576868}
  \pgfmathsetmacro{\baseone}{-0.313879*pi}

  \pgfmathsetmacro{\atwo}{0.450555}
  \pgfmathsetmacro{\rtwo}{0.618443}
  \pgfmathsetmacro{\basetwo}{0.586117*pi}

  \pgfmathsetmacro{\phione}{\baseone - pi/4}      
  \pgfmathsetmacro{\phitwo}{\basetwo - 3*pi/4}    

  \begin{axis}[
    width=11cm,
    height=7cm,
    xlabel={$\beta$ (rad)},
    ylabel={$\varphi$ (rad)},
    xmin=0, xmax=3.141593,
    grid=none,
    legend style={at={(1.00,0.5)}, anchor=east,font=\small},
  ]

    \addplot+[thick, smooth, mark=none, color=myPink, domain=0:3.141593,samples=400]
    {
      (
        atan2(
          \rone * sin(deg(\phione)) * sin(deg(2*x)), 
          \aone * cos(deg(2*x)) +
          \rone * cos(deg(\phione)) * sin(deg(2*x))  
        ) * pi/180
        + (x > pi/2 ? -2*pi : 0)   
      )
      + x
    };
    \addlegendentry{LCP theory}

    \addplot+[thick, smooth, mark=none, color=myPurple, domain=0:3.141593,samples=400]
    {
      (
        atan2(
          \rtwo * sin(deg(\phitwo)) * sin(deg(2*x)), 
          \atwo * cos(deg(2*x)) +
          \rtwo * cos(deg(\phitwo)) * sin(deg(2*x))  
        ) * pi/180
        + (x > pi/2 ? -2*pi : 0)
      )
      + 3*x
    };
    \addlegendentry{RCP theory}

    \addplot+[thick,dashed, mark=none, color=myPin,]
    coordinates {
      (0.000000,0.000000)
      (0.087266,-0.139747)
      (0.174533,-0.293233)
      (0.261799,-0.438661)
      (0.349066,-0.576565)
      (0.436332,-0.694624)
      (0.523599,-0.796218)
      (0.610865,-0.871572)
      (0.698132,-0.933199)
      (0.785398,-0.985768)
      (0.872665,-1.029046)
      (0.959931,-1.070389)
      (1.047198,-1.109042)
      (1.134464,-1.155596)
      (1.221730,-1.208448)
      (1.308997,-1.270934)
      (1.396263,-1.348600)
      (1.483530,-1.450693)
      (1.570796,-1.572624)
      (1.658063,-1.712711)
      (1.745329,-1.861973)
      (1.832596,-2.010571)
      (1.919862,-2.148436)
      (2.007129,-2.270044)
      (2.094395,-2.366878)
      (2.181662,-2.439260)
      (2.268928,-2.505305)
      (2.356194,-2.557166)
      (2.443461,-2.600027)
      (2.530727,-2.642666)
      (2.617994,-2.682035)
      (2.705260,-2.727542)
      (2.792527,-2.781340)
      (2.879793,-2.844263)
      (2.967060,-2.845487)
      (3.054326,-3.022920)
      (3.141593,-3.143477)
    };
    \addlegendentry{LCP modelling}

    \addplot+[thick, line width=2pt, dotted, mark=none, color=myPurpl,]
    coordinates {
      (0.000000,0.000000)
      (0.087266,0.177780)
      (0.174533,0.372020)
      (0.261799,0.579996)
      (0.349066,0.792436)
      (0.436332,1.009374)
      (0.523599,1.225403)
      (0.610865,1.440109)
      (0.698132,1.646111)
      (0.785398,1.842263)
      (0.872665,2.018774)
      (0.959931,2.158047)
      (1.047198,2.221419)
      (1.134464,2.140199)
      (1.221730,1.800220)
      (1.308997,1.463046)
      (1.396263,1.369541)
      (1.483530,1.431035)
      (1.570796,1.569407)
      (1.658063,1.747455)
      (1.745329,1.944729)
      (1.832596,2.151364)
      (1.919862,2.364614)
      (2.007129,2.580963)
      (2.094395,2.796094)
      (2.181662,3.010838)
      (2.268928,3.216042)
      (2.356194,3.412002)
      (2.443461,3.588233)
      (2.530727,3.729361)
      (2.617994,3.796696)
      (2.705260,3.703797)
      (2.792527,3.364334)
      (2.879793,3.034769)
      (2.967060,2.938219)
      (3.054326,3.002865)
      (3.141593,3.141009)
    };
    \addlegendentry{RCP modelling}

  \end{axis}
\end{tikzpicture}
    \caption{Comparison between the simulated results for the nanoparticle and the theoretical predictions obtained using Eq.~\eqref{eq:alles_ist_hier}.
    The amplitude and phase values at $0$ and $\pi/4$ were taken directly from the simulation results to determine $a$ and $b$.
    }
    \label{fig:Erleuchtung}
\end{figure}
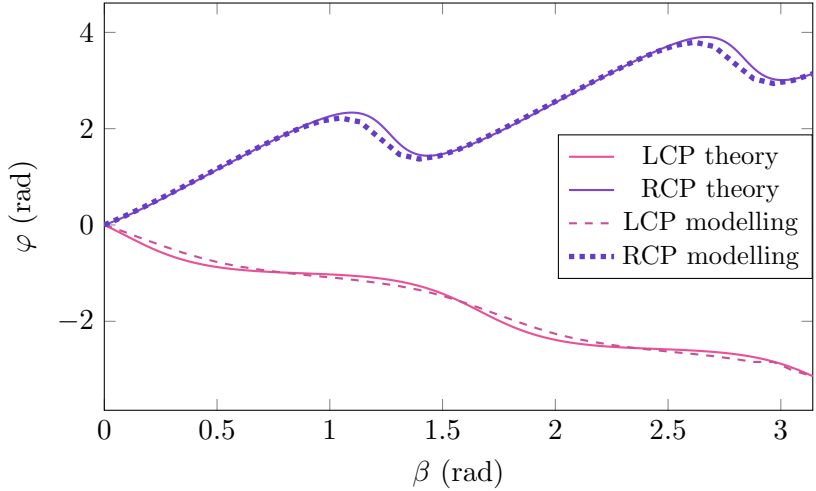
Let us consider the dependence of the SH amplitude shown in Fig.~1b of the main text.
Note that the following considerations are only valid for AlGaAs; however, the approach can be applied to other materials.
In~\cite{Nikitina2023}, it is shown that the $\chi^{(2)}$ tensor of AlGaAs transforms under rotations around the $z$-axis in the same way as $\sin(2\phi)$, where $\phi$ is the azimuthal angle of the cylindrical coordinate system.
This means that the tensor $\chi^{(2)\prime}$ rotated by an angle $\beta$ can be expressed as follows~\cite{video2}:
\begin{equation}
\label{eq:tensrot}
    \chi^{(2)}\prime= \cos(2\beta)\chi^{(2)}_1 +    \sin(2\beta)\chi^{(2)}_2 
\end{equation}
where $\chi^{(2)}_2$ is obtained from $\chi^{(2)}_1$ by a rotation of $\pi/4$, and we assume that $\chi^{(2)}_1$ corresponds to $x \parallel [100]$.
Let us first assume that the crystalline lattice, rather than the nanostructure itself, is rotated.
To describe the intensity, this assumption works, because changing the coordinate system to one aligned with the structure only alters the phase.
Due to linearity in the sense described in~\cite{menshikov2025near,Nikitina2024}, the SH electric field can be represented as a sum of the fields generated by the two terms in Eq.~\eqref{eq:tensrot}.
The two orientations of the tensor correspond to configurations 1 and 2 in Fig.~\ref{fgr:lattice}.
The second-harmonic field can then be written in the form
\begin{equation}
    \vb E^{2\omega}= \cos(2\beta)\vb E_1 +    \sin(2\beta)\vb E_2 
\end{equation}
where $\vb E_1$ and $\vb E_2$ are the fields corresponding to positions 1 and 2 of Fig.~\ref{fgr:lattice}, but with different phases due to the rotation of the coordinate system.
Each of them contains both polarization components (as well as terms with other TAM projections).
The coefficient preceding a given polarization component is expressed by an arbitrary complex number determined by the resonances of the structure, rather than by symmetry.
We denote these coefficients in both terms as $a$ and $b$.
Accordingly, the intensity dependence on $\beta$ is given by
\begin{equation}
    I(\beta) \sim \left|a\cos(2\beta)\vb  +  b \sin(2\beta)\right|^2
\end{equation}
where $a$ and $b$ are complex numbers that can be determined through numerical modelling.
This formula is applicable to AlGaAs nanostructures, but not to our metasurfaces, because in the case of a metasurface we rotate the unit cell in the modelling rather than the metasurface as a whole; as a result, different rotation angles lead to different modal content.
If, instead, the metasurface were rotated as a whole relative to the crystalline lattice, this approach would also be applicable.
To obtain the phase, one must be more precise.
Let us repeat the rotation procedure from the previous section while accounting for the time shifts.
This yields the same phase terms, $-m^{2\omega}\beta + 2m^{\omega}\beta$, but we must also rotate the tensor according to Eq.~\eqref{eq:tensrot}.
The final phase can then be expressed as
\begin{equation}
     \varphi(\beta)= \arg(a\cos(2\beta)\vb  +  b \sin(2\beta)) - m^{2\omega}\beta + 2m^\omega\beta
     \label{eq:alles_ist_hier}
\end{equation}
where $a$ and $b$ are, in principle, different for the generated LCP and RCP components.
This expression describes the dependence shown in Fig.~1c of the main text (see Fig.~\ref{fig:Erleuchtung}).





\section{Pancharatnam phase and holonomy}
\begin{figure}[h!]
\begin{center}
  \includegraphics[width=0.65\textwidth]{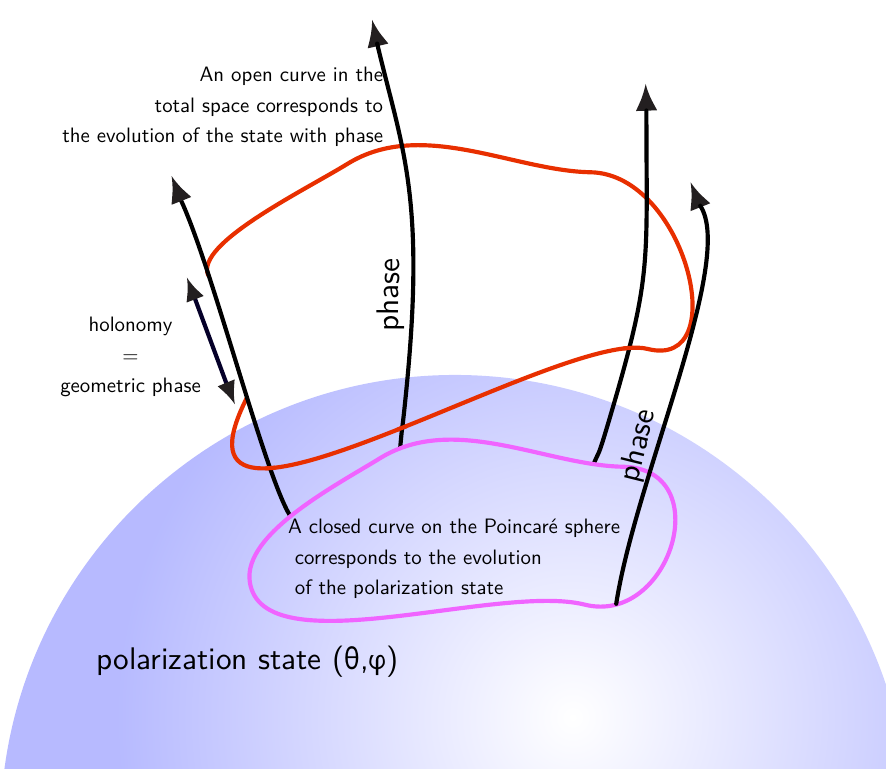}
  \caption{Geometry of the geometric phase on the example of the Pancharatnam phase.
  \textbf{Locally}, the space of a state with phase resembles the direct product of a point on the Poincaré sphere and a phase coordinate, in analogy with viewing three-dimensional space as a plane times a line.
  However, this structure is globally non-trivial.
  When the polarization state undergoes a cyclic evolution, the corresponding curve in the total space is not closed.
  Note that the path in this total space is determined by the choice of a connection.
}
  \label{fgr:geom}
\end{center}
\end{figure}
Pancharatnam–Berry phases, as well as geometric phases in solid-state physics, are naturally described as holonomies of a $U(1)$ connection~\cite{PhysRevLett.51.2167,Nakahara,RevModPhys.94.031001,Cayssol2021Apr}.
In the Pancharatnam case, the polarization state traces a closed contour on the Poincaré sphere.
In the associated total space $S^{3}$, the state including its phase evolves according to a specific rule, for instance one imposed by a waveplate in the classical Pancharatnam setup.
When the full complex state is considered, the corresponding trajectory in the total space $S^{3}$ of normalized Jones vectors in $\mathbb{C}^{2}$ is generally not closed; the resulting mismatch after one cycle is exactly the holonomy~\cite{Nakahara,RevModPhys.94.031001} (see Fig.~\ref{fgr:geom}).
All other standard geometric phases follow the same logic: one considers a closed contour in the base space whose lift to the total space is not closed, and the phase is given by the corresponding holonomy.
One can introduce a connection and its curvature, and, in principle, even Chern classes.

In the nonlinear case, such a construction is not straightforward.
We deal with the generation of a state at frequency~$2\omega$ rather than with an evolution on the Poincaré sphere, and the observed process cannot be associated with any closed loop in the space of polarization states.
No natural connection or holonomy emerges, and we therefore find it difficult to interpret this phase as geometric in the strict sense.

On the other hand, one may adopt a broader convention, frequently used in the community, according to which any phase that depends on some ``geometric'' degree of freedom (e.g., orientation of a scatterer) is called geometric. 
Under such an extended definition, many additional phases would fall into the category of geometric. 

However, if one extends the definition even further, the distinction becomes ambiguous. 
For instance, consider a vortex beam: its azimuthal phase profile $\exp(\iu m\varphi)$ varies solely because of geometry in the most literal sense: it depends on the coordinate on the transverse plane.
But it has no relation to any holonomy or to parallel transport.
According to the loose usage of the term, one might be tempted to call this phase geometric as well, yet it does not correspond to a geometric phase in the holonomic sense. 
This illustrates that ``depending on geometry'' is probably not a sufficient criterion.


\bibliographystyle{unsrtnat}  
\bibliography{bibliography} 